\documentclass[onecolumn,journal]{IEEEtran}
\usepackage{setspace}
\usepackage{amsmath,amsfonts}
\usepackage{algorithmic}
\usepackage{algorithm}
\usepackage{array}
\usepackage{textcomp}
\usepackage{stfloats}
\usepackage{url}
\usepackage{verbatim}
\usepackage{graphicx}
\usepackage{cite}
\usepackage[justification=centering]{caption}
\usepackage{subcaption}
\usepackage{longtable}

\usepackage{authblk}

\IEEEoverridecommandlockouts

\begin{document}

\title{Open RAN: Evolution of Architecture, Deployment Aspects, and Future Directions}

\author[1,2]{Prabhu Kaliyammal Thiruvasagam}
\author[1]{Chandrasekar T}
\author[1]{Vinay Venkataram}
\author[1]{Vivek Raja Ilangovan}
\author[1]{Maneesha Perapalla}
\author[1]{Rajisha Payyanur}
\author[1]{Senthilnathan M D}
\author[1]{Vishal Kumar}
\author[1]{Kokila J}
\affil[1]{NEC Corporation India, Chennai, India 600096}
\affil[2]{Indian Institute of Technology Madras, Chennai, India 600036}

\maketitle
\doublespacing

\begin{abstract}

The Open Radio Access Network (Open RAN) aims to enable disaggregated, virtualized, programmable, and data-driven intelligent network with open interfaces to support various real-time and non-real-time applications for different classes of users and multiple industry verticals in beyond 5G and 6G networks while providing interoperability among multi-vendor network functions and components. In this article, we first discuss the evolution of RAN and then the O-RAN Alliance standardization activities and objectives to provide a comprehensive overview of O-RAN from a standardization point of view. Then, we discuss the O-RAN security aspects, use cases, deployment aspects, and open source projects and related activities in other forums. Finally, we summarize the open issues, challenges, and future research directions to explore further for in-depth study and analysis.  

\end{abstract}


\section{Introduction}

It is estimated that around 13.2 billion Internet of Things (IoT) devices are already connected to the mobile networks till November 2022 and the number will increase to approximately 35 billion by 2028 \cite{Ericsson_MR}. As the number of IoT devices connected to the mobile networks are rapidly increasing, the mobile data traffic will increase in multi-fold between 2022 and 2028,  with compound annual growth rate of 24\% and almost doubled in last two years \cite{Ericsson_MR}. Furthermore, beyond 5G and 6G networks are envisioned to support new case cases and act as drivers to achieve the UN's Sustainable Development Goals by 2030 \cite{UN_SDG} \cite{NGMN_6G}. Hence, network operators are under pressure to support new services and use cases while reducing the overall Capital Expenditures (CapEx) and Operational Expenditures (OpEx) in a highly competitive mobile services markets. Recent technologies such as Network Functions Virtualization (NFV) and Software-Defined Networking (SDN) have changed the way of designing, deploying, and operating/managing network functions at the core network by leveraging the features of virtualization, cloud computing, and decoupling control plane from user plane \cite{NFV_SDN} \cite{NFV_5G} \cite{3gpp_23.501}. NFV and SDN have enabled to design a more agile and less expensive core networks using virtual machines, containers,  and programmable controllers. However, it is estimated that 65-70\% of the CapEx and OpEx is spent only for the Radio Access Network (RAN) part alone in an entire total cost of ownership \cite{O-RAN_WP1} \cite{O-RAN_PW}. Because the traditional RAN, which consists of Radio Unit (RU) and Baseband Unit (BBU), is implemented as all-in-one architecture and each RAN operates independently \cite{O-RAN_survey2}. Hence, the traditional RAN is also referred to as Distributed RAN (D-RAN).

In all-in-one traditional RAN deployment, RU and BBU are integrated as a single unit and thus it is easy to implement. However, the traditional hardware based RAN network functions are \cite{O-RAN_survey1}: i) proprietary and closed, ii) expensive and designed as a single unit, and thus often seen as a black-box by network operators, iii) not quickly scalable and agile, iv) not reconfigurable for network operators unique needs without the help of vendors, v) not easily integrable and interoperable with other vendor equipments, and v) not designed to enable automation through intelligent networks. To overcome these limitations and reduce the overall expenditures, the traditional RAN has to be re-designed to improve the efficiency of RAN deployments and operations of RAN network functions.

Over the last decade, many research and standardization activities (e.g., C-RAN, vRAN, and xRAN) have been carried out to re-design RAN deployments in order to reduce both CapEx and OpEx \cite{O-RAN_PW} \cite{O-RAN_survey2} \cite{O-RAN_survey1}. They are majorly focused on disaggregating the traditional RAN components RU and BBU into two or more separate units (BBU can be further split into Distributed Units (DUs) and Central Unit (CU)) and share a pool of BBU processing units in a central physical location with multiple RUs. A new interface called the Fronthaul (FH) interface is used to connect and communicate between the pool of BBU processing units and RUs. This approach has helped to reduce the overall expenditures significantly by leveraging the features of cloud computing and standardize some interfaces (e.g., FH interface) \cite{O-RAN_survey2} \cite{O-RAN_survey1}. However, still there are problems such as the software and some of the interfaces are remain either proprietary or closed that prevent interoperability and a multi-vendor ecosystem, vendor lock-in and monopoly prevent competitive and vibrant supplier system which could reduce the network equipment cost, and new management mechanisms are required to handle the increased complexity to support new 5G services and use cases~\cite{O-RAN_PW}~\cite{O-RAN_survey2}. To overcome these problems, RAN interfaces and network functions need to be made open to enable interoperability through multi-vendor deployments, healthy competition, and vibrant supplier ecosystem by allowing smaller vendors participation \cite{O-RAN_PW} \cite{O-RAN_survey2} \cite{O-RAN_survey1}. In addition, RAN virtualization and network automation through intelligent controllers can be considered to efficiently manage and orchestrate network services in an intelligent manner with reduced cost~\cite{O-RAN_survey2}~\cite{Mavenir_WP1}.

Open RAN is the movement in mobile industry to disaggregate RAN components and create open interfaces between them. Two industry groups are majorly leading the Open RAN movement. The first one is Telecom Infra Project (TIP) which formed in 2016 \cite{TIP}. TIP mainly focuses on development and deployment of open, disaggregated, and multi-vendor interoperable RAN solutions through the OpenRAN project group \cite{TIP_OpenRAN}. TIP does not create any specifications on its own, but uses specifications to create multi-vendor interoperable solutions for network operators and service providers through different project groups. The other one is O-RAN Alliance which formed in 2018 by merging two related organizations C-RAN Alliance and xRAN Forum~\cite{O-RAN_PW}~\cite{O-RAN_web1}. The O-RAN Alliance leads the mobile industry towards open interfaces, interoperable RAN ecosystem, RAN virtualization, and data-driven RAN intelligence by creating specifications and supporting open source projects \cite{O-RAN_web1}. TIP and O-RAN Alliance have signed a liaison agreement for sharing information, referencing, and validation activities \cite{TIP_OpenRAN}. Both TIP and O-RAN Alliance encourage the use of open source software and Commercial-Off-The-Shelf~(COTS) open white box hardware for RAN deployments \cite{O-RAN_WP1} \cite{TIP_OpenRAN}. The core principles of the O-RAN Alliance are open interfaces, cloudification, and automation through closed-loop control \cite{O-RAN_web1}. The O-RAN Alliance leverages the features of NFV, SDN, and Artificial Intelligence/Machine Learning (AI/ML) to realize its vision \cite{O-RAN_WP1} \cite{O-RAN_WP2} \cite{O-RAN_WP3}. It is estimated that NFV, SDN, and AI/ML based network deployment and management can save about 50\% of total cost of ownership \cite{O-RAN_survey2}. 

In the literature, current challenges faced by telecom industry players, state-of-the-art standardization activities, and future research directions considering the emerging technologies are not covered sufficiently. This paper, to fill that gap, provides a comprehensive overview of evolution of RAN architecture, disaggregated RAN functions and interfaces between them, O-RAN Alliance standardization activities (including O-RAN architecture, open interfaces, intelligent controllers for orchestration and automation of RAN, virtualization and cloud infrastructure, use cases, security aspects, and base station white box deployment), related open source projects, and system related open issues, challenges, and future research directions to explore further for in-depth study and analysis.

The rest of the paper is organized as follows: Section II provides background details about the evolution of RAN and O-RAN Alliance. Section III discusses the O-RAN Alliance reference architecture, main network functions, and protocol stack. Section IV explains about the need for open interfaces, 3GPP defined interfaces, and O-RAN Alliance defined open interfaces in the O-RAN reference architecture. Section V explains about functionalities of RAN Intelligent Controllers for RAN optimization and automation. Section VI discusses virtualization and edge computing enabled NFV infrastructure. Section VII discusses network slicing and RAN slicing optimization related aspects and Section VIII discusses O-RAN security related aspects. Section IX discusses a set of use cases enabled by O-RAN network functions and nodes. Section X discusses about deployment aspects and related open source projects. Section XI discusses open issues, challenges, and future research directions. Finally, Section XII concludes the paper.

\section{Background}

\subsection{Evolution of RAN}
In mobile communication networks, the Radio Access Network (RAN) plays an important role to connect User Equipments~(UEs) with the public/private Core Network (CN) to register and access services by routing control plane and user plane data. The architecture of RAN has been evolved in each generation of the mobile communication networks to support new services and accommodate ever increasing users to access services seamlessly. In the 2nd Generation (2G) networks, RAN (also known as base transceiver station) was designed to support mainly voice services through circuit-switched network and controlled by base station controllers.  In the 3rd Generation (3G) networks, RAN (also known as Node Base station (NB)) was designed to support both voice and data services through circuit-switched and packet-switched networks and controlled by radio network controllers. In the 4th Generation (4G) networks, RAN was designed to support high data rate services through only packet-switched, Internet Protocol based architecture, network. To simplify the architecture and operations, RAN and its controller are integrated in 4G and called as evolved NB (eNB). 

The 5th Generation (5G) and beyond networks are envisioned to provide services not only to human beings, but also to multiple industry verticals such as transportation, agriculture, healthcare, energy, manufacturing, gaming, entertainment, and public safety services \cite{NGMN1_5G} \cite{NGMN2_5G}. 5G services are broadly classified into three categories \cite{IMT_2020}: i) enhanced Mobile Broadband (eMBB) which supports high data rate services such as ultra-high definition video broadcasting and online gaming, ii) Ultra-Reliable and Low Latency Communications (URLLC) which supports time sensitive and high reliable services such as remote surgery and autonomous vehicle driving, and iii) massive Machine Type Communications (mMTC) which supports to provide connectivity to billions of IoT devices. Compare to previous generation of networks, 5G network has stringent service requirements such as latency in the order of milliseconds, ultra-reliable networks, connectivity for billions of IoT devices, and 100x network energy efficiency compare to 4G \cite{IMT_2020} \cite{5g_req}.  

Network operators are transforming their networks to support new 5G based services and business use cases, handle the ever increasing number of connected terminals and mobile data traffic, and to optimize the available network resources to manage the load and offer diverse services with satisfiable Quality of Service/Experience (QoS/QoE). In particular, network operators leverage the features of NFV, SDN, Edge Computing, and Satellite Networks to meet the service requirements of the users/verticals while aiming to reduce their overall CapEx and OpEx \cite{NGMN2_5G}. However, it is estimated that around 65-70\% of the total cost of ownership is dedicated only for RAN deployments, operations, and maintenance \cite{O-RAN_WP1} \cite{O-RAN_PW}. Therefore, network operators are focusing to reduce the CapEx and OpEx of RAN by redesigning RAN in a flexible manner such that different deployment options and new services can be supported.   

In general, RAN consists of RU and BBU for radio signal transmission/reception and radio resource management, respectively. In traditional RAN or D-RAN (e.g., NB and eNB), both RU and BBU are integrated in all-in-one architecture for easy implementation and thus each D-RAN operates independently. As the number of UEs attempt to attach to the network increases, additional D-RANs need to be deployed in a dense manner to accommodate additional requests. In this case, network operators have to spend more money to increase the capacity and maintain the QoS/QoE. This is because the traditional RAN networks are closed, proprietary, and designed as a single entity. Thus, CapEx and OpEx are major issues when it comes to base station deployments and operations \cite{O-RAN_WP1} \cite{O-RAN_PW} \cite{O-RAN_survey2}. 

To reduce the overall expenditures, first Cloud RAN (C-RAN) concept was proposed \cite{C-RAN}. C-RAN exploited the idea of RAN node resource sharing. The C-RAN concept is also referred to as Centralized RAN. The idea is to split RU and BBU of RAN into two separate units and process signalling information of multiple RUs in a central pool of BBUs. In C-RAN architecture, the RU is also referred to Remote Radio Unit/Head (RRU/RRH) as it is located at a distant place in a cell tower compared to a pool of BBUs in a single physical location \cite{C-RAN}. Here, the pool of BBUs are placed in a cloud data center. C-RAN supports centralized processing, collaborative radio, real-time radio networking, and scalability. A new FH interface is used as transport network to connect and communicate between the BBU pool and a set of RUs~\cite{C-RAN2}. The industry standardized Common Public Radio Interface (CPRI) can be used for FH implementation \cite{C-RAN3}. The C-RAN architecture based deployments has reduced energy consumption, CapEx, and OpEx. However, there are limitations such as a single point of failure for BBU, large FH overhead and throughput limitations, security issues, and a proprietary interface based deployment results in vendor lock-in.   

Virtualized RAN (vRAN) also follows a similar approach as C-RAN and employs virtualization additionally. vRAN replaces proprietary and expensive hardware with COTS hardware and decouples the software from hardware by applying the principles of NFV \cite{vRAN2}. Network functions can be deployed in a flexible manner using vRAN on top of virtual machines or containers and there is no hardware dependency. Hence, vRAN enhances flexibility, scalability, and total cost of ownership saving. However, coordination among RAN nodes to distribute network resources for different services based on the service requirements is challenging, complexity of the network management has increased significantly, and a proprietary interface based deployment results in vendor lock-in.

Extensible RAN (xRAN) follows a similar approach as vRAN and has done the following in addition by applying the principles of NFV and SDN: i) decouples the RAN control plane from the user plane, ii) builds a modular BBU software stack that operates on COTS hardware, and iii) defines open north- and south-bound interfaces \cite{XRAN}. xRAN has proposed a standardized FH, and promoted the idea of managing RAN nodes through RAN controller with data analytics platform and standardizing the interfaces between BBU and RAN controller \cite{Mavenir_WP1}. xRAN can support to meet the 5G performance requirements. To unify the efforts, the O-RAN Alliance formed by merging C-RAN Alliance and xRAN Forum in 2018. The O-RAN Alliance leads the Open RAN movement to realize the open, disaggregated, virtualized, and automated RAN deployments. Table \ref{tab:table-1} compares different characteristics of the RAN types and its evolution.

\begin{table}[]
	\centering
	\caption{Comparison of characteristics of RAN types and evolution.}
	\label{tab:table-1}
	
	\begin{tabular}{|c|c|c|c|c|c|c|c|}
		\hline
		\textbf{Types} & \textbf{\begin{tabular}[c]{@{}c@{}}RU and BBU \\ Separation\end{tabular}} & \textbf{\begin{tabular}[c]{@{}c@{}}BBU at \\ Cloud\end{tabular}} & \textbf{Virtualization} & \textbf{\begin{tabular}[c]{@{}c@{}}Control and User\\  Plane Separation\end{tabular}} & \textbf{\begin{tabular}[c]{@{}c@{}}Standardized\\ Interfaces\end{tabular}} & \textbf{\begin{tabular}[c]{@{}c@{}}Multi-vendor\\ Support\end{tabular}} & \textbf{Automation} \\ \hline
		D-RAN & No                                                               & No                                                      & No             & No                                                                           & NA                                                                & No                                                             & No         \\ \hline
		C-RAN & Yes                                                              & Yes                                                     & No             & No                                                                           & Partially Yes                                                     & Partially Yes                                                  & No         \\ \hline
		vRAN  & Yes                                                              & Yes                                                     & Yes            & No                                                                           & Partially Yes                                                     & Partially Yes                                                  & No         \\ \hline
		xRAN  & Yes                                                              & Yes                                                     & Yes            & Yes                                                                          & Mostly Yes                                                        & Mostly Yes                                                     & Mostly Yes \\ \hline
		O-RAN & Yes                                                              & Yes                                                     & Yes            & Yes                                                                          & Yes                                                               & Yes                                                            & Yes        \\ \hline
	\end{tabular}
\end{table}

\subsection{3GPP and 5G}
\label{section:3GPP and 5G}

The 3rd Generation Partnership Project (3GPP) is a consortium formed in 1998 by seven regional telecommunications standard development organizations to develop global standards for 3G \cite{3GPP_Intro}. Later, the scope of 3GPP was extended and is responsible for developing standards for beyond 3G system such as 4G and 5G \cite{3GPP_Intro}. 3GPP standard specifications are structured in terms of Releases \cite{3GPP_Releases}. 3GPP specifications are primarily defined as three-stage process: i) stage 1 specifications define the service requirements from the user point of view, ii) stage 2 specifications define an architecture to support the service requirements defined in stage 1, and iii) stage 3 specifications define an implementation of the architecture defined in stage 2 by specifying protocols in details. 3GPP has been defined 5G System architecture for different scenarios and use cases since Release 15 in different phases \cite{3gpp_23.501}. The 5G System Phase 1 was specified in 3GPP Release 15, Phase 2 was specified in Release 16, and the capability of the 5G System was enhanced further in 3GPP Release 17. Currently, 3GPP Releases 18 and 19 are being specified as part of 5G Advanced and they are open. 

In 5G System, the RAN is called Next Generation RAN (NG-RAN) and the new air interface for 5G is called as New Radio (NR). 5G NR is a new radio access technology which is a physical connection method for radio based communication. Two frequency bands are defined for 5G NR: Frequency Range 1 (FR1) which includes sub-6 GHz frequency bands and Frequency Range 2 (FR2) which includes frequency bands from 24.25 GHz to 71 GHz \cite{3gpp_38.101-1} \cite{3gpp_38.101-2}. NG-RAN supports both 5G NR radio access and 4G Long Term Evolution (LTE) radio access \cite{3GPP_38.300}. An NG-RAN node (i.e., base station) can be either gNB (5G base station) which provides NR radio access towards UE or ng-eNB (evolved 4G base station that connected to 5G core) which provides LTE radio access towards UE \cite{3GPP_38.300}. The gNBs and ng-eNBs are interconnected via Xn interface. NG-RAN supports Multi-Radio Dual Connectivity (MR-DC) in which a UE can connect to two NG-RAN nodes (LTE-NR or NR-NR)~\cite{3GPP_38.300}~\cite{3GPP_37.340}.     

From deployment point of view, 3GPP 5G System supports two modes: Non-Standalone (NSA) and Standalone (SA). In NSA mode, LTE and NR are tightly integrated and connected to the existing 4G CN (Evolved Packet Core (EPC)), leveraging the MR-DC towards the UE. In MR-DC architecture, gNB and eNB provide radio resources concurrently towards the UE for high data rate. In NSA mode, eNB handles both control plane and user plane data and gNB handles only user plane data. In SA mode, gNB is connected to the 5GC directly and the MR-DC with two NG-RAN nodes (NR-NR) can be leveraged for high data rate. In SA mode, both gNBs handles control plane and user plane data. The NSA mode of deployment can support to provide eMBB class of services and it is expected that the SA mode of deployment can support URLLC and mMTC class of services with 5G Advanced system. 

\begin{figure*}[!t]
	\centering
	\begin{subfigure}{0.5\textwidth}
		\centering
		\includegraphics[scale=0.6]{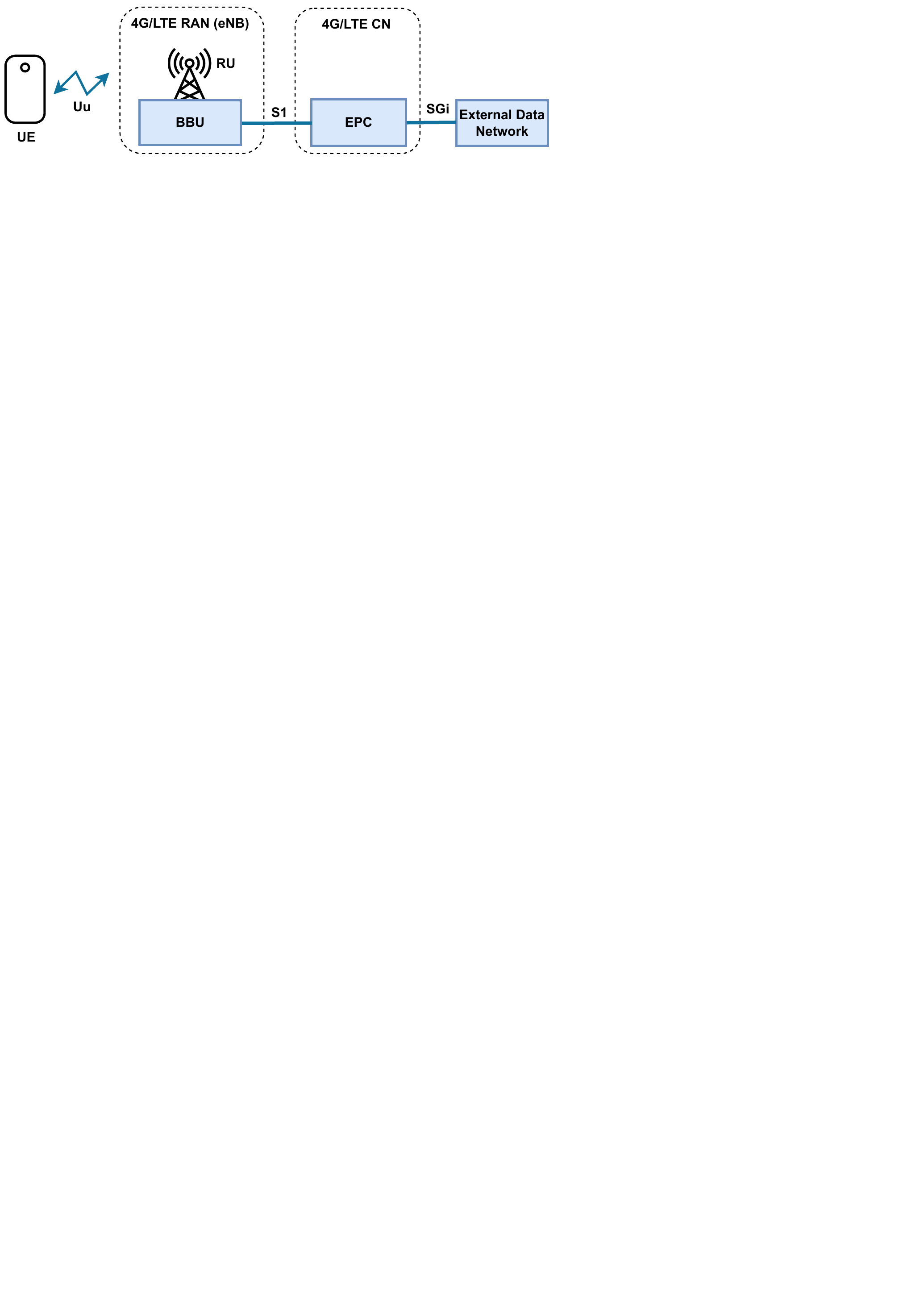}
		\vspace{-16cm}
		\caption{4G architecture}
		\label{fig:4G_arch}
	\end{subfigure} 
	\hspace{-1cm}
	\begin{subfigure}{0.5\textwidth}
		\centering
		\includegraphics[scale=0.45]{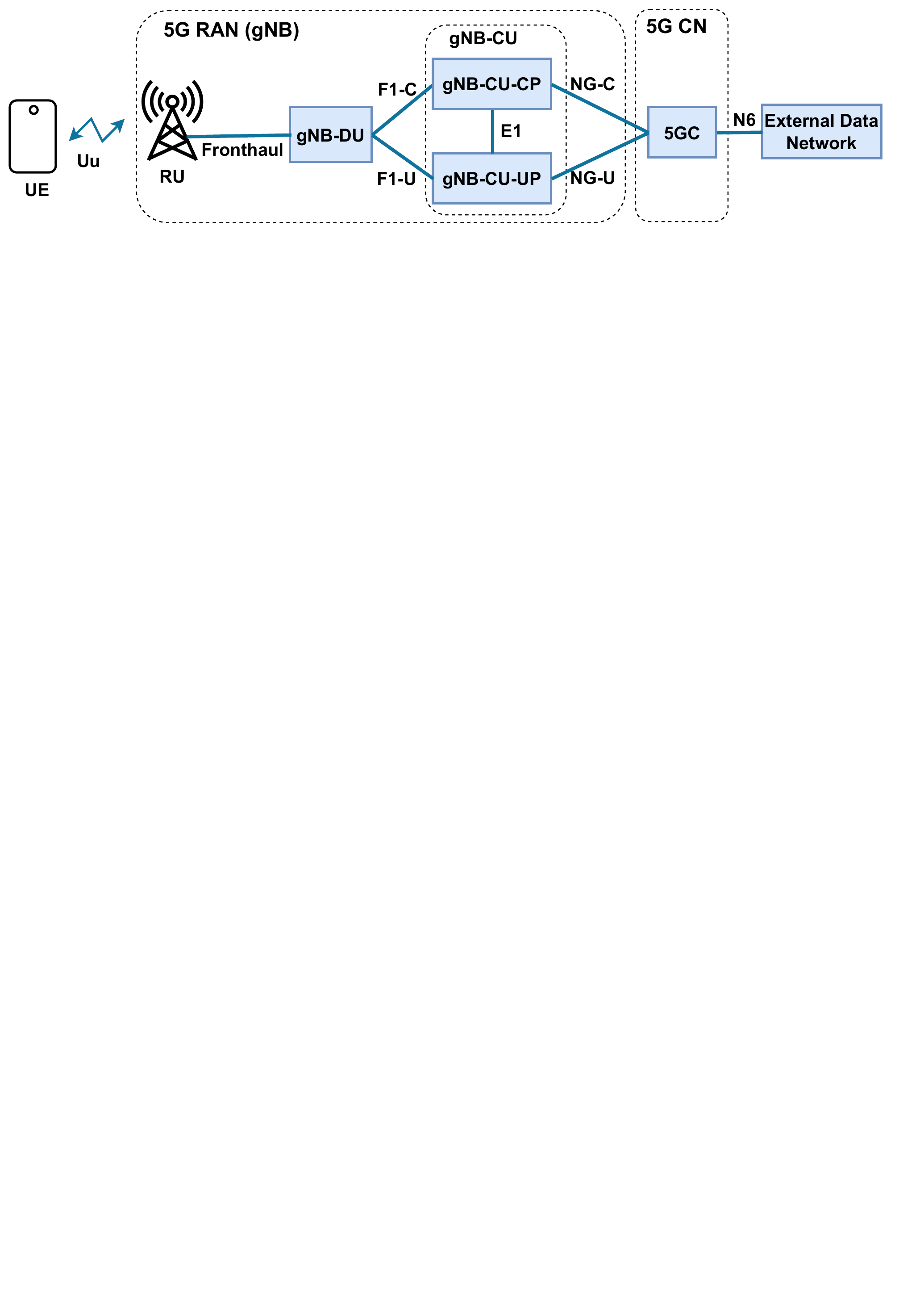}
		\vspace{-11.5cm}
		\caption{5G architecture}
		\label{fig:5G_arch}
	\end{subfigure}
	\caption{4G and 5G architectures.}
	\label{fig:4G_5G}
\end{figure*}

4G and 5G architectures are shown in Figure \ref{fig:4G_5G}. In Figure \ref{fig:4G_arch}, 4G architecture depicts the traditional RAN deployment. As shown in Figure \ref{fig:5G_arch}, 5G architecture supports RAN disaggregation and an NG-RAN node gNB may consist of a Central Unit (gNB-CU) and one or more Distributed Units (gNB-DU) \cite{3GPP_38.401}. A gNB-CU and gNB-DU units are connected via F1 interface. One gNB-DU is connected to only one gNB-CU. A gNB is connected to 5G Core (5GC) via NG interface. The interfaces Xn, F1, and NG are logical interfaces and they can be split further with respect to control plane and user plane data (e.g., F1-C and F1-U). The gNB-CU can be split into gNB-CU-CP and gNB-CU-UP with respect to control plane and user plane separation, and they are interconnected via the E1 interface. In end-to-end 5G System architecture, N2 interface is used in the place of NG-C and N3 interface is used in the place of NG-U \cite{3gpp_23.501}. There are many deployment options that can be considered by network operators based on the requirements of the users \cite{3GPP_38.801}. NG-RAN supports eight different functional split options between CU and DUs from higher layer split to lower layer split to realize the enhanced performance \cite{3GPP_38.801}.

5G NG-RAN supports disaggregation of RAN functionalities, flexible functional split options, and control plane and user plane separation for effective scaling. However, open interfaces and network functions, virtualization, optimization, and automation are not yet fully supported to offer various services based on the requirements of the users/industry verticals and enable interoperability \cite{O-RAN_PW} \cite{O-RAN_survey2}. Moreover, management of mobile networks is increasingly complex due to the advent of 5G and densification of heterogeneous disaggregated networks operating in multi-band frequency ranges to support multiple industry verticals and massive IoT connectivity. To overcome these limitations and issues, the mobile networks need to be virtualized, software-driven, flexible, scalable, intelligent, reconfigurable, interoperable, and energy efficient \cite{O-RAN_WP1}. O-RAN Alliance is working towards to fill this gap through open interfaces, intelligence, and virtualized edge computing infrastructure.

\begin{table}[]
	\centering
	\caption{O-RAN Alliance groups and their objectives.}
	\label{tab:WGs}
	\begin{tabular}{|l|l|lll}
		\cline{1-2}
		\multicolumn{1}{|c|}{\textbf{Groups}} & \multicolumn{1}{c|}{\textbf{Objectives}}                                                                                                                    &  &  &  \\ \cline{1-2}
		\multicolumn{1}{|c|}{WG1}             & Focuses on O-RAN reference architecture and use cases                                                                                                                                                         &  &  &  \\ \cline{1-2}
		\multicolumn{1}{|c|}{WG2}             & \begin{tabular}[c]{@{}l@{}}Focuses on the non-RT RIC and its relevant open interfaces for non-RT \\ automation and optimization of RAN Radio Resource Management\end{tabular} &  &  &  \\ \cline{1-2}
		\multicolumn{1}{|c|}{WG3}             & \begin{tabular}[c]{@{}l@{}}Focuses on the near-RT RIC and its relevant open interfaces for near-RT \\ control and optimization of RAN elements and resources\end{tabular}      &  &  &  \\ \cline{1-2}
		\multicolumn{1}{|c|}{WG4}             & Focuses on the open Fronthaul interfaces                                                                                                                                                                      &  &  &  \\ \cline{1-2}
		\multicolumn{1}{|c|}{WG5}             & \begin{tabular}[c]{@{}l@{}}Focuses on 3GPP defined NG-RAN interfaces for enabling fully interoperable \\ multi-vendor ecosystem\end{tabular}                                                                  &  &  &  \\ \cline{1-2}
		\multicolumn{1}{|c|}{WG6}             & Focuses on RAN cloudification and orchestration                                                                                                                                                               &  &  &  \\ \cline{1-2}
		\multicolumn{1}{|c|}{WG7}             & Focuses on the white box hardware design for open base station deployments                                                                                                                                    &  &  &  \\ \cline{1-2}
		\multicolumn{1}{|c|}{WG8}             & Focuses on stack reference design based on the NR protocol stack                                                                                                                                              &  &  &  \\ \cline{1-2}
		\multicolumn{1}{|c|}{WG9}             & Focuses on the new transport network based on fronthaul, midhaul, and backhaul                                                                                                                                &  &  &  \\ \cline{1-2}
		\multicolumn{1}{|c|}{WG10}            & Focuses on OAM and its relevant interfaces                                                                                                                                                                    &  &  &  \\ \cline{1-2}
		\multicolumn{1}{|c|}{WG11}            & Focuses on security aspects of O-RAN reference architecture                                                                                                                                                                                   &  &  &  \\ \cline{1-2}
		\multicolumn{1}{|c|}{FG1}             & Focuses on O-RAN standardization development strategies                                                                                                                                                       &  &  &  \\ \cline{1-2}
		\multicolumn{1}{|c|}{FG2}             & Focuses on open source projects and related activities                                                                                                                                                        &  &  &  \\ \cline{1-2}
		\multicolumn{1}{|c|}{FG3}             & Focuses on testing and integration                                                                                                                                                                            &  &  &  \\ \cline{1-2}
		\multicolumn{1}{|c|}{RG}              & Focuses on next generation open and intelligent RAN principles                                                                                                                                                  &  &  &  \\ \cline{1-2}
	\end{tabular}
\end{table}

\subsection{O-RAN Alliance}
The O-RAN Alliance formed in 2018 by a group of leading network operators and vendors. The O-RAN Alliance aims to transform RAN towards open, intelligent, virtualized, fully interoperable, and autonomous RAN, while improving the performance, cost efficiency, and agility \cite{O-RAN_web1}. The core principles of O-RAN Alliance foster open interfaces and functions, multi-vendor interoperable RAN ecosystem, faster innovation, supporting virtualization, leveraging open source software, running on COTS white box hardware in cloud infrastructure, and automation of RAN using AI and ML \cite{O-RAN_web1}. 
The O-RAN Alliance specifies its principles on top of 3GPP 4G and 5G RANs (LTE and NR). Particularly, the O-RAN Alliance takes the disaggregated NG-RAN (shown in Figure \ref{fig:5G_arch}) as base and additionally introduces two RAN Intelligent Controllers (RICs), open interfaces, and NFV based cloud infrastructure for enabling openness, automation, and cost-efficiency. RICs aid for automation and optimization of RAN elements and resources in Real-Time (RT) and non-RT to support diverse services on-demand without violating the service requirements. The RICs are connected with the disaggregated NG-RAN components via open interfaces for enabling fully interoperable multi-vendor ecosystem. The O-RAN reference architecture is shown in Figure \ref{fig:O-RAN_arch}.

The O-RAN Alliance is actively involved in three main streams: Specification Work, O-RAN Software Community (SC), and Testing and Integration. The O-RAN  specification work has been divided into technical Work Groups (WG), Focus Groups (FG), and Research Groups (RG) to improve the efficiency of RAN deployments (e.g., energy- and spectral-efficient) and operations of mobile networks (e.g., cost-effective and agile) \cite{O-RAN_web2}. The O-RAN Alliance groups and their activities are listed in Table \ref{tab:WGs}.

\section{O-RAN Architecture}

O-RAN architecture has been built upon the foundation set forth by 3GPP 5G System. The O-RAN Alliance includes new RAN functions, open, and interoperable interfaces. The O-RAN Alliance's approach to RAN architecture is to provide additional benefits to network operators such as minimizing deployment cost (e.g., using white box COTS servers) and increased supply chain diversity (e.g., more vendors and vibrant supply chain ecosystem) while handling complexity through intelligence and enabling interoperability among different vendors. The main changes in O-RAN architecture in comparison with 3GPP NG-RAN architecture are: introduction of new nodes such as RICs and service management entities, introduction of new open interfaces, and the cloud infrastructure.

\begin{figure}[!t]
	\centering
	\includegraphics[scale=0.5]{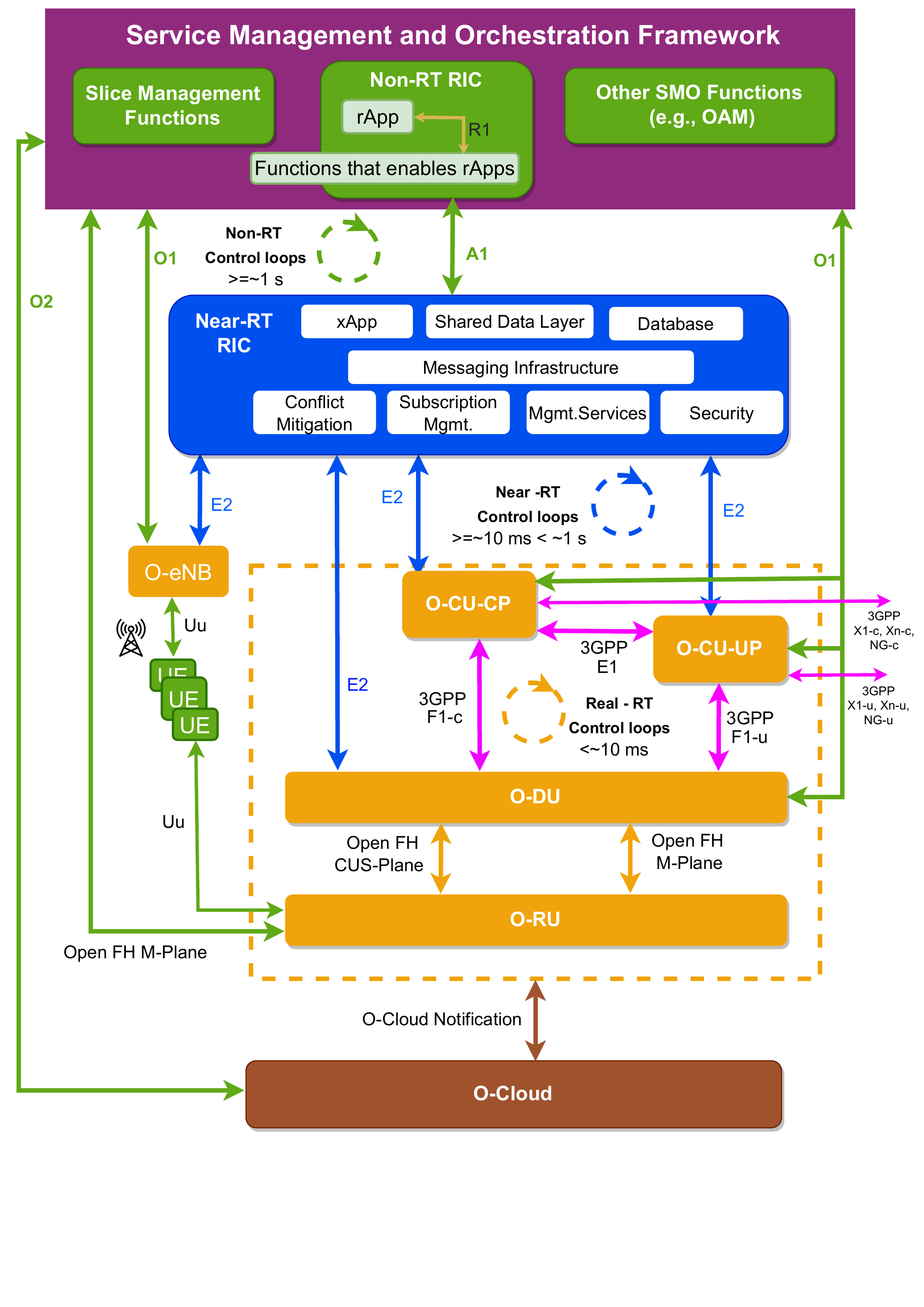}
	\vspace{-1.8cm}
	\hspace{3cm}\caption{O-RAN reference architecture.}	
	\label{fig:O-RAN_arch}
\end{figure}

Figure \ref{fig:O-RAN_arch} shows the O-RAN reference architecture with RAN nodes/functions and interfaces, which provides the foundation for building virtualized RAN on open hardware in cloud infrastructure with intelligent radio controllers powered by AI/ML~\cite{O-RAN_web1}~\cite{W1a}. The O-RAN architecture consists of various physical and logical components which are connected through different interfaces. The O-RAN reference architecture is being developed by O-RAN Alliance WG1 \cite{W1a}, which mainly consists of: i) Service Management and Orchestration (SMO) Framework, ii) RICs and closed-loop control, iii) O-RAN enabled CU (O-CU), iv) O-RAN enabled DU (O-DU), v) O-RAN enabled RU (O-RU), vi) O-RAN enabled Cloud Infrastructure (O-Cloud), vii) O-RAN enabled eNB (O-eNB), and viii) UEs. The radio side of the O-RAN reference architecture includes near-RT RIC, O-CU-CP, O-CU-UP, O-DU, and O-RU functions, whereas the management side includes SMO Framework containing a Non-RT-RIC and other management functions. The main entities of the O-RAN reference architecture are briefly described below. 

\subsection{SMO Framework}

The 5G RAN as envisioned by O-RAN is designed to be highly flexible, quickly scalable, and multi-vendor interoperable for different deployment models. Thus, the management and automation aspects of the RAN attain paramount importance. The SMO framework introduced in the O-RAN architecture (shown in Figure \ref{fig:O-RAN_arch}) is a collection of integrated functions and services that is in charge of handling all the management, orchestration, and automation aspects of various O-RAN components. In principle, the SMO is similar to the Management and Orchestration (MANO) entity in NFV reference architecture \cite{NFV_arch} \cite{NFV-MANO_arch}. SMO operates on Service Based Architecture (SBA) principles to provide and consume services (e.g., authentication, authorization, service registration and discovery, data management, trained model sharing etc.) using standardized service-based interfaces which enable interoperability within SMO functions \cite{W1a}.
SMO provides Fault, Configuration, Performance, Authentication, and Security (FCAPS) management functions to O-RAN nodes/components and O-Cloud via different interfaces (including O1, O2, A1, and Open FH interfaces). SMO also supports RAN optimization through non-RT RIC, O-Cloud management via O2 interface, orchestration, and workflow management \cite{W1a}.

\subsection{RICs and Closed-Loop Control}

With the emergence of data-driven intelligent AI/ML technologies, the O-RAN Alliance aims to leverage these in order to automate RAN deployment and operations. The O-RAN Alliance introduces two flavors of the RIC as software-defined components that are responsible for controlling and optimizing the O-RAN elements and resources. The first one is non-Real-Time RIC which deals with functionality and operations at a larger time scale (more than 1s). The other one is near-Real-Time RIC which deals with functionality and operations at a smaller time scale and close to real-time (10ms to 1s). The RICs connect to each other via the A1 interface and to other O-RAN components via the E2 interface.
The O-RAN architecture incorporates the concept of control loops, which can be defined as closed-loop autonomous action and feedback loops intended for network optimization by providing real-time intelligence and analytics \cite{Control_Loops}. The O-RAN reference architecture supports three types of control loops: non-RT loops, near RT loops, and RT loops. The Non-RT RIC runs non-RT control loop and typical execution time is more than one second. The Near-RT RIC runs near-RT control loop and typical execution time is less than one second and more than ten milliseconds. The E2 nodes (O-RAN nodes that terminate E2 interfaces) run RT control loop and typical execution time is less than ten milliseconds. Multiple Control loops can run simultaneously and interact with each other depending on the use case requirements. The interaction between different control loops and network functions for various use cases are specified in~\cite{W1c}~\cite{W1h}. RICs use data collected from various O-RAN functions (e.g., number of users, users' mobility data and resource usage) to create an abstract and centralized view of the network and can use AI and ML algorithms and training models to automate the RAN operations (e.g., RAN slicing, handovers, scheduling policies, and managing conflicts) through closed-loop controls.

\subsection{O-RU}
O-RU is a radio unit located in cell sites which transmits, receives, and processes the radio signals at the physical layer of the network. Through the Open FH interface, the radio signals are communicated between the O-RU and the O-DU in the O-RAN reference architecture. As per the functional split option 7.2x, the O-RU implements the PHY-Low and Radio Frequency (RF) processing functions as shown in Figure \ref{fig:O-RAN_stack}. The  O-RU terminates the Open FH interface for communication towards other O-RAN components as listed in Table \ref{tab:interfaces}.

\subsection{O-DU}

\begin{figure}[!t]
	\centering
	\includegraphics[scale=0.7]{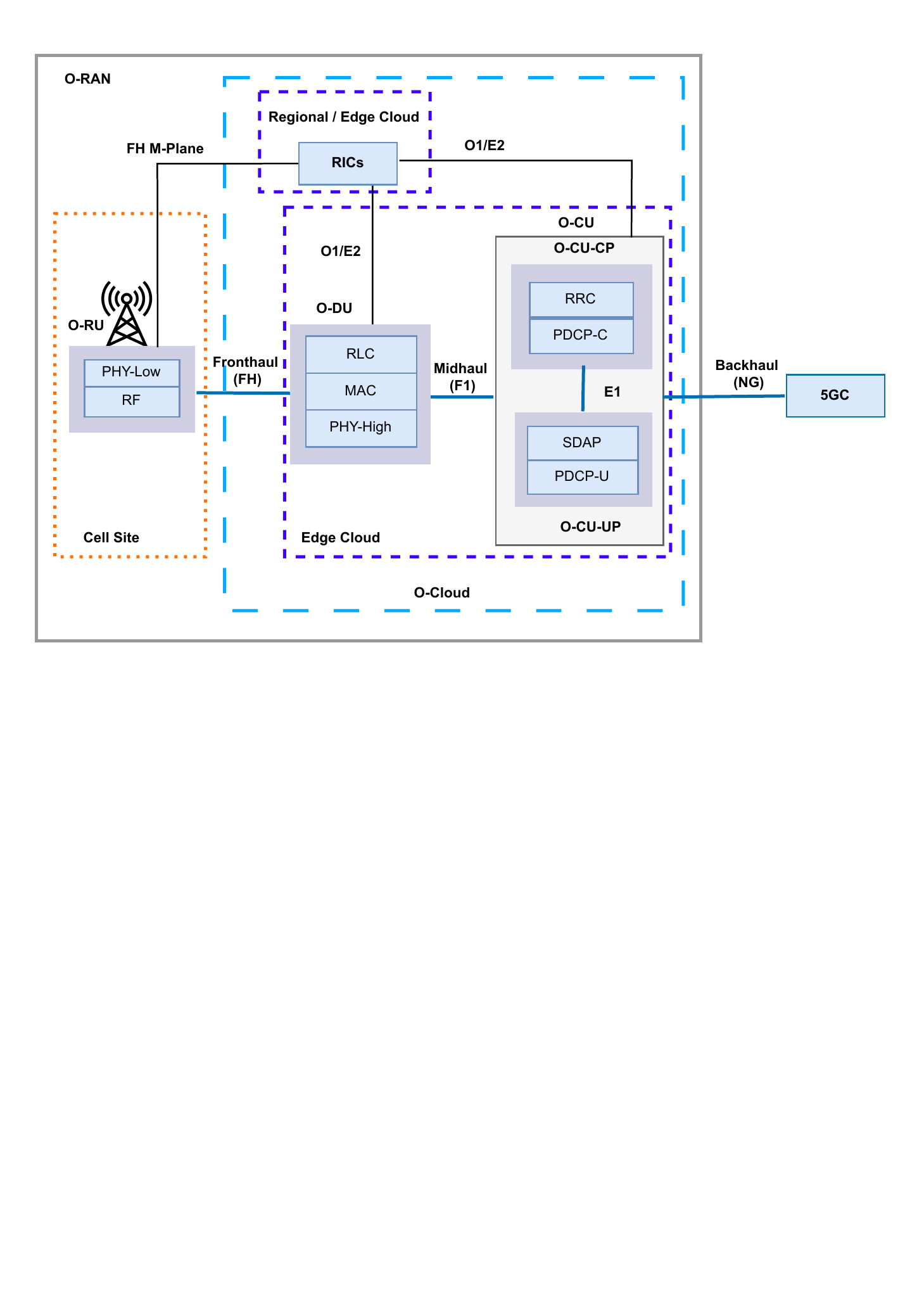}
	\vspace{-7.3cm}
	\hspace{3cm}\caption{O-RAN architecture with a functional split.}	
	\label{fig:O-RAN_stack}
\end{figure}

O-DU is a logical node that hosts part of the protocol functions of O-RAN flexibly in order to support diverse use cases. For instance as per the functional split option 7.2x, PHY-High, Medium Access Control (MAC), and Radio Link Control (RLC) protocols of the base station are hosted by O-DU as shown in Figure \ref{fig:O-RAN_stack} \cite{W1a}. The functionality of the O-DU can be split into two logical nodes: one is implementing the PHY-High function and the other is implementing the MAC and RLC functions on the received radio data\cite{O-RAN_survey1}. These two logical nodes can be used to implement the Small Cell Forum's standard interface Functional Application Platform Interface (FAPI) \cite{W1a} \cite{SCF} \cite{SCF1}.

\subsection{O-CU}
O-CU is a logical node that implements a set of base station protocols such as Radio Resource Control (RRC), Packet Data Convergence Protocol (PDCP), and Service Data Adaptation Protocol (SDAP) \cite{W1a}. O-RAN architecture leverages the SDN concept of decoupling the control plane from the user plane and applies it to split the CU into CU-CP and CU-UP. This methodology of splitting the CU into separate logical nodes is inherited from 3GPP and uses the E1 interface, which allows easy scaling and cost effective solutions for UP. Decoupling also enables advanced control functionality for better RRM via data-driven optimization, closed-loop control, and automation using advanced AI/ML tools. 
As shown in Figure \ref{fig:O-RAN_stack}, O-CU-CP runs RRC and PDCP-C and O-CU-UP runs SDAP and PDCP-U \cite{W8a}. The network functions O-DU, O-CU-CP, O-CU-UP run in the edge cloud and RICs can run either in edge cloud or regional cloud of the O-Cloud infrastructure as shown in Figure \ref{fig:O-RAN_stack}.

\subsection{O-Cloud}
O-Cloud is a cloud computing platform that consists of a collection of physical infrastructure nodes to host O-RAN functions (e.g., near-RT RIC, O-CU-CP, O-CU-UP, and O-DU), support software components (e.g., Operating Systems, Virtual Machines, Containers, Network Function Images, and Templates), and support management and orchestration related functions (e.g., Infrastructure Managers, Network Function Managers, Service Orchestration, Resource Orchestration, and RAN Slice Managers). O-Cloud Notification interface is used to notify O-Cloud related information to O-RAN network functions. 
O-Cloud can host RAN functions in edge cloud and regional cloud as shown in Figure \ref{fig:O-RAN_stack}. The functions O-DU, O-CU-CP, and O-CU-UP are placed in edge cloud to meet the latency requirements. Near-RT RIC can be placed either in edge cloud or regional cloud, whereas non-RT RIC is placed in regional cloud. There can be more than one edge clouds and regional clouds that are managed and orchestrated by O-Cloud and SMO. 

\subsection{O-eNB}
O-eNB is an O-RAN enabled eNB or ng-eNB which supports O-DU and O-RU functions with an Open FH interface between them \cite{W1a}. O-eNB also supports E2 interface related functions and operations, and it is connected to near-RT RIC via E2 interface \cite{W1a}. 

\subsection{UEs}
UEs are attached to O-RAN via Uu interface and can connect to O-eNB and O-gNB to access services simultaneously using MR-DC \cite{3GPP_38.300} \cite{3GPP_37.340}. UEs can be smartphones used by human beings and/or static and mobile IoT devices (e.g., sensors, machines, and vehicles) to support diverse vertical services.

\begin{table}[!t]
	\centering
	\caption{O-RAN interfaces.}
	\label{tab:interfaces}
\begin{tabular}{|c|l|c|}
\hline
\textbf{Interfaces} &
  \multicolumn{1}{c|}{\textbf{Description}} &
  \textbf{\begin{tabular}[c]{@{}c@{}}Managing \\ Authority\end{tabular}} \\ \hline
A1 &
  \begin{tabular}[c]{@{}l@{}}A1 is the interface between the Non-RT RIC function in SMO and \\ the Near-RT RIC function\end{tabular} &
  \begin{tabular}[c]{@{}c@{}}O-RAN\\ Alliance\end{tabular} \\ \hline
O1 &
  \begin{tabular}[c]{@{}l@{}}The O1 interface is between O-RAN Managed Element (SMO) and \\ the management entity (Radio side nodes)\end{tabular} &
  \begin{tabular}[c]{@{}c@{}}O-RAN\\ Alliance\end{tabular} \\ \hline
O2 &
  \begin{tabular}[c]{@{}l@{}}The O2 interface is between the SMO and O-Cloud to provide O-Cloud\\ platform resources and workload management\end{tabular} &
  \begin{tabular}[c]{@{}c@{}}O-RAN\\ Alliance\end{tabular} \\ \hline
E2 &
  E2 is a logical interface connecting the near-RT RIC with an E2 Node &
  3GPP \\ \hline
E1 &
  This interface is between O-CU-CP and O-CU-UP functions &
  \begin{tabular}[c]{@{}c@{}}O-RAN\\ Alliance\end{tabular} \\ \hline
F1-c &
  This F1-c interface is between the O-CU-CP and the O-DU functions &
  3GPP \\ \hline
F1-u &
  The F1-u interface is between the O-CU-UP and the O-DU functions &
  3GPP \\ \hline
NG-c &
  The NG-c interface is between O-CU-CP and the 5GC &
  3GPP \\ \hline
NG-u &
  The NG-u interface is between the O-CU-UP and the 5GC &
  3GPP \\ \hline
X2-c &
  \begin{tabular}[c]{@{}l@{}}The X2-c interface is for transmitting control plane information\\ between eNBs or between eNB and en-gNB in EN-DC\end{tabular} &
  3GPP \\ \hline
X2-u &
  \begin{tabular}[c]{@{}l@{}}The X2-u interface is for transmitting user plane information\\ between eNBs or between eNB and en-gNB in EN-DC\end{tabular} &
  3GPP \\ \hline
Xn-c &
  \begin{tabular}[c]{@{}l@{}}Xn-c interface is for transmitting control plane information \\ between gNBs, ng-eNBs or between   ng-eNB and gNB\end{tabular} &
  3GPP \\ \hline
Xn-u &
  \begin{tabular}[c]{@{}l@{}}The Xn-u interface is for transmitting user plane information\\ between gNBs, ng-eNBs or between ng-eNB and gNB\end{tabular} &
  3GPP \\ \hline
Uu &
  The Uu interface is between the UE to eNB/gNB &
  3GPP \\ \hline
R1 &
  R1 interface is between rApps and functions that enable rApps in SMO &
   \begin{tabular}[c]{@{}c@{}}O-RAN\\ Alliance\end{tabular} \\ \hline
Y1 &
  Y1 interface is between Y1 consumer and Near-RT RIC &
   \begin{tabular}[c]{@{}c@{}}O-RAN\\ Alliance\end{tabular} \\ \hline
\begin{tabular}[c]{@{}c@{}}O-Cloud \\ Notification\end{tabular} &
  \begin{tabular}[c]{@{}l@{}}This interface is within O-Cloud where event consumer\\ subscribes to events/status of O-Cloud\end{tabular} &
  \begin{tabular}[c]{@{}c@{}}O-RAN\\ Alliance\end{tabular} \\ \hline
CTI &
  CTI is interface between the O-DU and TN &
  \begin{tabular}[c]{@{}c@{}}O-RAN\\ Alliance\end{tabular} \\ \hline
\begin{tabular}[c]{@{}c@{}}Open \\ Fronthaul\end{tabular} &
  \begin{tabular}[c]{@{}l@{}}This interface is between O-DU and O-RU, and it is\\ further split into CUS-plane for and M-plane\end{tabular} &
  \begin{tabular}[c]{@{}c@{}}O-RAN\\ Alliance\end{tabular} \\ \hline
\end{tabular}
\end{table}

\section{3GPP NG-RAN INTERFACES AND O-RAN OPEN INTERFACES}
As discussed in Section \ref{section:3GPP and 5G}, 3GPP is a strong proponent for disaggregation of the RAN in 5G deployments and thus introduced various new network functions and interfaces thereby providing benefits related to lower deployment cost (using COTS servers) and increased supply chain diversity (more vendors and choices for operators) \cite{OpenInterfaces1}. However, this would result in an increased network complexity and integration costs. A good example of this problem can be seen at the interface between the DU and the RU. This interface is known as the FH interface and operates over the CPRI protocol. 3GPP leaves the implementation of this FH interface and CPRI protocol up to to the vendors choice \cite{OpenInterfaces2}, thereby limiting the interoperability among vendors. Another limitation of the 3GPP defined interfaces is the X2 interface, which is used for inter-connectivity between eNBs, but unfortunately this was left optional and upto implemention of vendors. This becomes an issue when deploying 3GPP 5G NSA architecture, where this X2 interface is used for connectivity between existing eNBs and newly deployed gNBs. This ties down operators to reuse the existing 4G vendors in the space.

A primary goal of the O-RAN Alliance is to resolve these issues regarding interoperability and multi-vendor deployments. Apart from 3GPP interfaces, the O-RAN Alliance has introduced a new set of interfaces that enable connectivity to the newly introduced components in the O-RAN architecture like Near-RT RIC, Non-RT RIC and also improves upon existing interfaces such as FH and X2 interface.

The new Open FH interface is introduced to enable open connectivity between various implementations of O-DU and O-RU. The O-RAN Alliance WG4 is tasked with developing and maintaining specifications for the Open FH interface. The Open FH interface is subdivided into Open FH CUS-Plane interface and Open FH M-Plane interface \cite{W4b}. The Open FH CUS-Plane interface get rids of the traditional vendor specific CPRI protocol and implements a newer and open standard called enhanced CPRI (eCPRI) \cite{W4b} \cite{eCPRI}.
The Open FH M-Plane interface facilitates connectivity between O-RU and it's managing entity (either O-DU or SMO). It specifically deals with the initialization, configuration and management aspects of the O-RU. Depending on the location of the management component, the Open FH M-Plane can be classified as hierarchical model (where O-RU is managed entirely by an O-DU) or hybrid model (where O-RU is managed by the SMO) \cite{W4g}.

The O-RAN Alliance has also introduced a new set of open interfaces (e.g., O1, O2, A1, E2, and O-Cloud Notification) to leverage virtualization, cloud computing features, and network automation and service management using different types of RICs. The O-RAN Alliance has specifically created a new O1 interface to facilitate a common and open approach to implement management functionality between various managed elements and any management entity \cite{W1e} \cite{W1d}. The O-RAN Alliance WG5 is tasked with developing and maintaining the specifications for O1 interface terminating at O-DU or O-CU. It creates the O1 specification for O-CU used over the interface linking the O-CU with the SMO \cite{W5e}. It also creates the O1 specification for O-DU used over the interface linking the O-DU with other management plane entities (that may include the O-CU as well as SMO) \cite{w5g}. The O-RAN Alliance WG5 also focuses on existing 3GPP defined NG-RAN interfaces (e.g., X2, Xn, E1, F1, and NG) and complements the existing standards by defining further O-RAN specifications related to C-Plane functions, U-Plane functions, and transport network \cite{W5b} \cite{W5c} \cite{W5j}. The aim is to promote improved openness and fully operable multi-vendor deployments using existing 3GPP defined interfaces. The set of interfaces defined in the O-RAN reference architecture are listed in Table \ref{tab:interfaces}.

\section{RICs, Automation, and Optimization of the RAN}

Human operators cannot handle the increased complexity in traditional way of deploying, optimizing, and operating the next generation mobile networks \cite{XAI}. Thus, AI and ML features can be leveraged to automate the operations of network functions and reduce operational expenditures. Using AI and ML algorithms and models, intelligence can be embedded in every layer of the RAN architecture to enable dynamic local radio resource allocation on demand and optimize network-wide efficiency \cite{O-RAN_WP1}. A combination of O-RAN open interfaces, RAN virtualization, and AI-powered closed-loop control can enable multi-vendor deployments, automation of network operations, and optimization of heterogeneous resources \cite{O-RAN_WP1}.    

The O-RAN reference architecture enhances the Radio Resource Management (RRM) functionalities with embedded intelligence by introducing the Non-RT RIC and Near-RT RIC through the A1 and E2 interfaces, respectively. RT analytics that drive embedded AI and ML backend modules will empower network intelligence \cite{O-RAN_WP1}.  

\subsection{Non-RT RIC}
Non-RT RIC is a logical function resides in SMO and handles the data carried across the A1 interface and non-RT RIC applications. The primary goal of non-RT RIC is to support non-real-time intelligent RRM, higher layer procedure optimization, policy optimization in RAN, and providing AI/ML models to near-RT RIC and other RAN functions. The A1 interface supports communication and information exchange between non-RT RIC and near-RT RIC. The key objective of A1 interface is to support policy-based guidance of near-RT RIC functions, transmission of enrichment information in support of AI/ML models into near-RT RIC, and basic feedback mechanisms from near-RT RIC. Non-RT RIC manages non-RT control functionalities, which may take more than one second time ($>$ 1s). The A1 interface is between Non-RT RIC in the SMO and O-gNB/O-eNB containing Near-RT RIC as shown in Figure \ref{fig:O-RAN_arch}. The Non-RT RIC shares the messages generated from AI-enabled policies, ML-based training models, and RT control functions with the Near-RT RIC for runtime execution through A1 interface. Also, network management applications running in Non-RT RIC can utilize the data received from O-DU and O-CU in a standardized format and the processed/analysed data can be shared to Near-RT RIC to make decisions quickly. The Non-RT RIC core algorithms developed and managed by network operators to modify RAN operations based on the individual policies and objectives for different deployment models. Non-RT RIC applications are called as rApps which are modular applications and leverage the functionality exposed via the non-RT RIC framework/SMO Framework to provide value added services related to RAN operation and optimization. The scope of rApps includes, but is not limited to, RRM, data analytics, and providing enrichment information.
The Non-RT RIC and A1 interface related aspects are being specified by the O-RAN Alliance WG2~\cite{W2b} .

\subsection{Near-RT RIC}

The O-RAN reference architecture enhances the RRM to support network intelligence via RICs. The primary goal for the near-RT RIC architecture is to specify the near-RT control functionalities and interfaces with CU/DU. Near-RT RIC is a logical function which supports to control RAN elements and optimize RAN resources in near-RT via fine-grained data collection and actions through E2 interface \cite{W1a}. Near-RT RIC handles near-RT control functionalities, which may take less than one second time ($<$ 1s), such as radio configuration management, mobility management, RAN slicing, QoS management, interference management, load balancing, connectivity and seamless handover management, enhanced/novel RRM functionalities, and resource block management. Near-RT RIC supports a robust, secure, and scalable platform that allows for flexible on-boarding of 3rd party applications (i.e., the application can be independent of the near-RT RIC). An application designed to run on the near-RT RIC is called as xApp which may consist of one or more microservices to consume and provide data. The xApps and related functions can leverage the database called Radio Network Information Base (R-NIB) which collects data from the underlying network via E2 interface and Non-RT RIC via A1 interface. From functionality point of view, xApp shall be able to receive event-triggered information on RAN information and time-varying network state and provide collected logging, tracing, and metrics information to Near-RT RIC. Near-RT RIC may include AI and ML workflow including model training, inferences, and updates \cite{W3a}. Various RAN measurements data are fed to near-RT RIC via E2 interface for better RRM. The E2 interface also carries the configuration commands directly from near-RT RIC to CU and DU. Near-RT RIC will execute new models (e.g., training models, policy decisions, and traffic and mobility predictions) received from non-RT RIC to change the functional behaviour of the network and its applications. The RRM functional allocation between the Near-RT RIC and the E2 Node is subject to the capability of the E2 Node exposed over the E2 interface by means of the E2 Service Model~\cite{W3g}. The Near-RT RIC and E2 interface related aspects are being specified by the O-RAN Alliance WG3 \cite{W3c}. 

\section{Virtualization and Cloud Infrastructure}

\subsection{NFV Infrastructure}

O-RAN network functions can be realized as Virtualized Network Functions (VNFs) and/or Physical Network Functions~(PNF). VNFs can run on Virtual Machines (VMs) and/or Containers on top of NFV Infrastructure (NFVI) in O-Cloud, and PNFs can be specially designed on a dedicated hardware to support certain features (e.g., reliability and security). NFVI platforms can be used to support various functional split options such as high layer split between PDCP and RLC and low layer split within PHY. These functions can be implemented as software modules on top of COTS hardware inside private or public cloud infrastructures. This leads to effectively utilize the available resources wherein multiple network functions can be virtualized and run on the same COTS hardware and reduces the overall CapEx and OpEx. The O-RAN framework follows similar principles of ETSI NFV which focuses on virtualization, softwarization, and cloudification \cite{NFV_5G}. The O-RAN Alliance WG6 specifies the cloud architecture and deployment scenarios to run O-RAN network functions on VMs and OS containers in~\cite{W6f}, similar to ETSI NFV framework. RAN cloudification is one of the fundamental key principles of the O-RAN architecture. The O-RAN Alliance WG6 also has specified a set of use cases regarding the deployment of O-RAN network functions on O-Clouds, as well as relevant functional and interface requirements between the SMO and O-Cloud \cite{W6h}. 

The primary goal of NFV is to disaggregate hardware and software parts of the network functions defined as part of O-RAN. As defined in \cite{W6f}, an NFV architecture would broadly be categorized into three layers: a bottom hardware layer (this maps to the ETSI NFVI hardware sublayer in the case of VM/container based deployment), a middle layer that includes cloud stack functions as well as Acceleration Abstraction Layer functions (this maps to the ETSI NFVI virtualization sublayer and Virtualized Infrastructure Manager (VIM) in the case of VM/container based deployment), and thirdly a top layer that supports the virtual RAN functions such as O-DU, O-CU-CP, O-CU-UP, and near-RT RIC.

The ETSI NFV specifications define the NFV MANO architectural framework which is a set of functional blocks and interfaces enabling the deployment and management of VM and container based VNFs \cite{NFV_arch} \cite{NFV-MANO_arch}. The management of virtualized and container based resources is performed according to a specified set of interfaces, models and APIs \cite{W6l}. The list of NFV MANO architectural framework functional blocks and their main responsibilities are \cite{NFV_arch} \cite{W6f}: i) VIM which is responsible for the management of virtualized resources (virtual compute, virtual network and virtual storage) in the NFVI and VM software images, ii) Container Infrastructure Service Management which is responsible for the management of container based resources and container workloads in Container Infrastructure Service clusters, iii) Container Image Registry which is responsible for the management of OS container software images, iv) VNF Manager which is responsible for the management of VNFs, including their lifecycle management, and provides corresponding management services based on a common set of interfaces and models regardless of the technology used for implementing the VNF, be it VM or container, and v) NFV Orchestrator which is responsible for the management of VNF Packages and other artifacts and the overall orchestration of NFVI resources across multiple sites (referred also as NFVI points-of-presence) and the lifecycle management of network services.

\subsection{Edge Computing}

Mobile network operators are investing in edge infrastructure, with market drivers including use cases from private cellular networks to edge robotics in industry, as well as disaggregated open networks. Currently, the applications are being processed in a way where less than 10\% of the traffic is actually processed outside of a traditional data center, which comes from the operator. But, the Mobile Edge Forum predicts that in the coming decade up to 75\% of the processing would happen outside of the traditional data centers due to the anticipated extrapolated growth in edge computing to various industry vertical services (e.g., Remote Healthcare Diagnostics, Industrial IoT, and online gaming) \cite{MEF}. With the introduction of the RIC function in O-RAN, the application hosting capability of operator networks favours edge computing and O-RAN network APIs. With edge computing the operators can host third-party RAN applications that are much closer to the consumer and have a per-user or per-device control mechanism in place. 5G use cases such as Traffic Steering, QoS Optimization, User Mobility Robustness among others can be supported with the help of edge computing to offer faster and higher quality "per-user" services as part of any dynamic network \cite{MEF}.

Multi-access/Mobile Edge Computing (MEC) enables the placement of applications close to the customer and the use of RAN contextual information. It supports seamless application mobility, provides service-oriented APIs for users’ location and radio conditions, and handles application-related traffic redirection. The MEC architecture consists of the MEC Platform~(MEP) that hosts MEC applications, the MEP Manager which is responsible for the management of platform and MEC applications life cycle, and the Virtualization Infrastructure and its manager \cite{MEC_GS003}. MEC leverages the features of NFV, SDN, network slicing, and AI/ML to offer services dynamically and improve the QoS and QoE \cite{MEC2}. 

The 5GC User Plane Function and 5G RAN (CU, DU, and RU) can be deployed at the network edge cloud for latency considerations and flexible execution of network functions based on the use case. MEC can be co-located with O-RAN to leverage the unique features available in both networks and reduce CapEx and OpEx. For instance, MEC hosts can be combined with the O-RAN near-RT RIC, the MEC based control plane services can be integrated with O-RAN services, MEC databases (about UE locations, cell performance, Radio Network Information Service) can be integrated with O-RAN databases, and the MEC Application Orchestrator can be used for xApps orchestration (including SON functions). The MEC application mobility mechanism can be reused in multi-near-RT RIC environment, solving an essential problem of the inter-near-RT RICs cooperation \cite{MEC2}.

\section{Network Slicing and RAN Slice Optimization}

\subsection{Network Slicing}
Network slicing is a key feature for 5G where multiple logical networks are created from a single physical infrastructure, with isolation of resources and optimized topology to serve a specific service category such as mMTC use cases which require low-power wide area communication, eMBB use cases that need high data rates, or URLLC use cases such as autonomous driving which demands low latency and high reliability \cite{3GPP_28.530}. Achieving E2E network slicing with predictable QoS is essential for a growing number of 5G services that depend on network slicing to operate at scale.

Network slicing overlays multiple virtual networks on top of a shared network domain, that is, a set of shared network and computing resources. 5G RAN slicing is part of an end-to-end (E2E) network slicing deployment for a 5G SA network. 
Network slicing enables service providers to maximize the use of network resources and service flexibility by leveraging the principles of 5G software and hardware disaggregation using NFV. This helps service providers to: i) create new revenue opportunities by lowering the barriers to trying out new service offerings, ii) increase service flexibility by enabling more kinds of services to be offered simultaneously, and iii) 	support rapid scaling and better time to market as all physical infrastructures are pooled as common shared resources \cite {slicing}.

The slicing framework can be broadly categorized into three distinct layers, namely: the Infrastructure Layer (IL), the Network Function Layer (NFL), and the Service Layer (SL), all three layers will be managed by a MANO entity. The IL broadly refers to the physical network infrastructure consisting of both RAN and CN. It also includes deployment, control and management of the infrastructure, the allocation of resources (computing, storage, network, radio) to slices, and the way that these resources are available to and managed by the higher layers. The NFL encapsulates all the operations that are related to the configuration and lifecycle management of the network functions that, after being optimally placed over the (virtual) infrastructure and chained together, offer an E2E service that meets certain constraints and requirements described in the service design of the network slice. The SL caters to the way that services should be described and how they should be mapped to the underlying network components, and the architecture of network slicing managers and orchestrators. An important element that distinguishes network slicing in the context of 5G from other forms of slicing that have been considered in the past (e.g., cloud computing) is its E2E nature and the requirement to express a service through a high-level description and to flexibly map it to the appropriate infrastructural elements and network functions. This observation regarding the operation of slicing in the context of 5G naturally leads to two new high-level concepts: i) a service layer that is directly linked to the business model behind the creation of a network slice and ii) network slice orchestration of a slice’s lifecycle management.

\subsection{Slice Monitoring and SLA Assurance}
Network slicing is the foundation for many use cases unique to 5G, ranging from private enterprise 5G network deployments with specific Service Level Agreement (SLA) requirements to low-latency applications like Virtual Reality (VR), Augmented Reality (AR), and Mixed Reality (MR). The SLAs are created between the mobile operator and the business customer based on the requirements and to ensure there are no violations. RAN slicing enables service differentiation handling on the RAN that allows for the effective use of dynamic radio resource partitioning, slice-aware QoS enforcement, and slice orchestration functionality for meeting SLA \cite{RAN_slicing}.

O-CU-CP and O-CU-UP should be slice aware and execute slice-specific resource allocation and isolation strategies. They initially configured through O1 based on slice-specific requirements and then dynamically updated through E2 via Near-RT RIC for various slicing use cases. These may generate and send specific performance measurements (PMs) through O1 and E2, where PMs can be used for slice performance monitoring and slice SLA assurance purposes. Network Slicing can be tailored to specific business requirements. O-RAN’s open interfaces and AI/ML based architecture will enable such challenging mechanisms to be implemented and help pave the way for operators to realize the features of network slicing in an efficient manner \cite{W1i}. 

\section{O-RAN Security Aspects}

The O-RAN Alliance WG11 has been focusing on security aspects of the O-RAN reference architecture network functions, interfaces, and cloud infrastructure. The objective is to build a secure, open, interoperable, and automated RAN. The focus of WG11 is to mitigating risk across O-RAN architecture in order to strengthen the security of Open RAN. The O-RAN Alliance has specified inherent security benefits of Open RAN \cite{W1a} (e.g., transparency and common control, interoperability of security protocols and security features, secure supply chain, and enhanced intelligence), possible threats \cite{W11d} (e.g., against architecture, cloud, supply chain, open source code, 5G radio networks, AI/ML system, and physical infrastructure), possible attack surface (e.g., added new functions and open interfaces, architecture modification, decoupling of software and hardware, 3rd party xApps and rApps, VMs, containers, and open  source software), risk assessment \cite{W11d}, security requirements \cite{W11b} (e.g., confidentiality, integrity, availability, authentication, authorization, and access control), security principles and controls~\cite{W1a}~\cite{W11a} (e.g., using secure protocols such as SSHv2, TLS 1.2 and 1.3, DTLS 1.2, IPSec, CMPv2, OAuth2.0, and NETCONF over Secure Transport on O-RAN interfaces), and security test \cite{W11c} (e.g., methodology, test cases, validation, and evaluation) to mitigate risk \cite{W1a}. Furthermore, the O-RAN Alliance WG11 is studying security aspects of different network functions~\cite{W11f}~\cite{W11e} (e.g., Non-RT RIC and Near-RT RIC) and cloud infrastructure \cite{W11g} (e.g., O-Cloud). 
	
It is expected that the security services offered by O-RAN should be at least with the same level as a 3GPP 5G NG-RAN~\cite{W1a}. Hence, the O-RAN Alliance is following the 3GPP security design principles and industry best practices to ensure the desired level of security expected by network operators and users. In addition, the O-RAN Alliance follows the principle of zero trust architecture which assumes no implicit trust of a user/asset based on location and ownership due to disaggregation of RAN and involvement of multiple industry players \cite{W1a} \cite{ZTA}.

\section{Use Cases}
The O-RAN Alliance WG1 has specified a set of use cases which employ AI/ML based methods to achieve the goal \cite{W1h} \cite{W1i}. The use cases can be broadly classified into three categories: user application related use cases, network resource optimization related use cases, and performance and QoS/QoE related use cases.  
\begin{itemize}
    \item Some of the user application related use cases are:
    \begin{itemize}
        \item \textit{Context-based dynamic handover management for Vehicle-to-Everything (V2X)} which improves the functionality of V2X applications by resolving handover related issues when vehicles move at high speed by using RICs and V2X Application Server.
        \item \textit{Radio resource allocation for UAV application scenario} which supports new applications using edge cloud and RICs.
        
    \end{itemize}
    \item Some of the network resource optimization related use cases are:
    \begin{itemize}
        \item \textit{RAN sharing} which increases network capacity and coverage while decreasing the cost of network implementation by using RICs.
        \item \textit{Dynamic Spectrum Sharing} which aids operators to dynamically share spectral resources already deployed between LTE and NR devices without compromising the QoE of the current 4G subscribers while providing the same level of coverage and essential QoS to NR devices by using RICs.
        \item \textit{Shared O-RU} which allows a controller (a NETCONF client) that determines and configures how the resources of a shared O-RU are partitioned between shared resource operators by using RICs.
    \end{itemize}
    \item Some of the performance and QoS/QoE realted use cases are:
    \begin{itemize}
        \item \textit{Energy saving} which turns off one or more carriers or the cell, energy saving can be accomplished over timescales of minutes, hours, and above when the cell load is minimal using RICs.
        \item \textit{Flight path based dynamic UAV radio resource allocation} which improves mobility performance and QoE using RICs.
        \item \textit{QoE Optimization} which aids for traffic recognition, QoE prediction and QoS enforcement decisions using RICs.
        \item \textit{QoS Based Resource Optimization} in which the network is equipped with features to maintain resource separation between slices and to keep track of whether each slice Service Level Specifications (SLS) are met using RICs. 
    \end{itemize}
\end{itemize}

\section{Deployment Aspects and Open Source Projects}
The O-RAN Alliance encourages the use of open source software and open white box hardware to reduce the deployment cost and develop Open RAN interoperable ecosystem that can be tested and integrated easily. The use of white box COTS hardware for designing software in a modular fashion enables to plan for scale out designs for capacity, availability, and reliability to support automation and optimization based on the service requirements. O-RAN white box hardware aspects for different base station deployment types are considered in WG7 to improve the performance, energy and spectral efficiency, and cost efficiency~\cite{W7a}. 

\subsection{Deployment Aspects}

Operator requirements for different deployment scenarios (e.g., Indoor Picocell, Outdoor Picocell, Outdoor Microcell, Integrated Access and Backhaul, and Outdoor Macrocell), use cases (e.g., eMBB and URLLC), and base station classes (e.g., local area, medium area, and wide area) for O-RAN white box hardware are being specified by WG7 \cite{W7a}. It also considers carrier frequency (e.g., FR1 and FR2), inter-site distance, and other base station related key attributes in the base station deployment scenarios. Key performance indicators such as peak data rate, peak spectral efficiency, bandwidth, latency, and mobility are considered to specify the requirements for both indoor and outdoor base station deployment scenarios~\cite{W7a}~\cite{W7h}~\cite{W7i}~\cite{W7j}. 
Base station architecture can be of three types for any deployment scenarios: i) split architecture (O-RU and O-DU are physically separated), ii) integrated architecture (O-RU and O-DU are implemented on one platform), and iii) all-in-one architecture (O-RU, O-DU, and O-CU are implemented on one platform) \cite{W7a}. 

As defined by 3GPP \cite{3GPP_38.801}, there are eight different FH split options. For white box hardware reference design architecture, the O-RAN Alliance supports three FH split options now: i) split option 6 (all PHY functions reside in O-RU and remaining RAN protocol functions reside in O-DU and O-CU), ii) split option 7.2x (RF and PHY-Low functions reside in O-RU and PHY-High, MAC, and RLC functions reside in O-DU, shown in Figure \ref{fig:O-RAN_stack}), and iii) split option 8 (Only RF function resides in O-RU and remaining PHY and other RAN protocol functions reside in O-DU and O-CU) \cite{W7c} \cite{W7d} \cite{W7e} \cite{W7f} \cite{W7g}. The O-CU shall be placed at a data centre and O-DU can be placed either at the data centre or at a cell site. Base station split architecture can be further divided into two types based on whether protocol translation is supported or not. FH Gateway (FHGW) can be used to interconnect O-RU/RU and O-DU, and FHGW supports protocol translation based on the FH split options \cite{W7b}. An optional FH Multiplexer (FHM) or a switch can be used in the place of FHGW to connect O-CU and O-DU with multiple or cascaded O-RU using FH interface. But, FHM or switch does not support protocol translation. 

O-RAN network functions and nodes (e.g., O-RU, FHM, FHGW, O-DU, O-CU-CP, O-CU-UP, Near-RT RIC, and SMO with Non-RT RIC) can be deployed in multiple ways depending on the operator's policies and use case requirements. The O-RAN Alliance WG1 has discussed the different implementation options of O-RAN functions and network elements in \cite{W1a}. For instance, all disaggregated nodes and functions can be deployed separately as defined in O-RAN reference architecture. It is also possible in the implementation to bundle some or all of these O-RAN nodes, and thus collapsing some of the internal interfaces such as F1-c, F1-u, E1 and E2 \cite{W1a}.

\subsection{Open Source Projects}

Currently, most of the telecommunication industry products use commercial software provided by software design companies. 
Open source software is generally not preferred for commercial products. However, in recent years there is an increase in acceptance and usage of open source software in software design companies and we are starting to see utilization of open source software in commercial Open RAN products. 
The development of open source software for O-RAN is being led by the O-RAN Software Community (SC). This a partnership between the O-RAN ALLIANCE and the Linux Foundation, with the goal of aiding in the development of open software for the RAN. The O-RAN SC is concentrated on aligning with the open architecture and requirements of the O-RAN Alliance to produce a solution suitable for commercial deployment. 
Some major internal projects led by the O-RAN SC are listed in Table \ref{tab:O-RAN SC Projects} \cite{O-RAN-SC}.

\begin{table}[]
\centering
\caption{O-RAN Software Community projects \cite{O-RAN-SC}.}
\label{tab:O-RAN SC Projects}
\begin{tabular}{|c|l|}
\hline
\textbf{Project Name} &
  \multicolumn{1}{c|}{\textbf{Description}} \\ \hline
RIC Applications &
  \begin{tabular}[c]{@{}l@{}}This project aims at developing sample xApps and platform \\ applications that can be used for integration, testing, and demos.\end{tabular} \\ \hline
Near Real Time RIC &
  \begin{tabular}[c]{@{}l@{}}This project aims at developing an initial RIC Platform that supports \\ xApps with limited support for O1, A1, and E2 interfaces.\end{tabular} \\ \hline
O-RAN Central Unit &
  \begin{tabular}[c]{@{}l@{}}This project aims an initial software deliverable with limited \\ functionality for the O-RAN Central Unit.\end{tabular} \\ \hline
\begin{tabular}[c]{@{}c@{}}O-RAN Distributed Unit \\ High Layers\end{tabular} &
  This project focuses on initial L2 functional blocks for the O-RAN DU. \\ \hline
\begin{tabular}[c]{@{}c@{}}O-RAN Distributed Unit \\ Low Layers\end{tabular} &
  This project focuses on initial L1 functional blocks for the O-RAN DU. \\ \hline
Simulations &
  \begin{tabular}[c]{@{}l@{}}This project aims at developing initial simulators used for testing \\ O-RAN NF interfaces.\end{tabular} \\ \hline
Infrastructure &
  \begin{tabular}[c]{@{}l@{}}This project focuses on developing the building blocks for infrastructure \\ to run O-RAN NF components.\end{tabular} \\ \hline
Non-Real Time RIC &
  \begin{tabular}[c]{@{}l@{}}This project aims at developing an initial Non-Real Time RIC Platform \\ that supports xApps with limited support for O1, A1, and E2 interfaces.\end{tabular} \\ \hline
\begin{tabular}[c]{@{}c@{}}Service Management and Orchestration \\ (SMO)\end{tabular} &
  \begin{tabular}[c]{@{}l@{}}The primary goal of the SMO project is to integrate different software \\ artifacts of existing open-source projects creating a fully functional \\ open-source SMO.\end{tabular} \\ \hline
AI/ML Framework &
  \begin{tabular}[c]{@{}l@{}}This project aims at creating an AI/ML workflow implementation for \\ O-RAN environment.\end{tabular} \\ \hline
\end{tabular}
\end{table}

Apart from the internal O-RAN SC led projects, there are multiple other 5G related software development by various external open source community projects.
Other major open source projects related to Open RAN are listed in Table \ref{tab:External Open-Source Projects}. Wide spread usage of open source RAN software enables the RAN software to be accessible to the majority of people, especially academics and research institutions, and can enable them to provide solutions to challenges faced by the industry. More research works using open source software can help to provide feedback on the limitations of the open source platforms and also on the O-RAN Alliance specifications. This creates a feedback loop where the solutions proposed by these open source communities and researchers can also be used for new feature developments in O-RAN specifications.

\begin{table}[]
\centering
\caption{External open source projects.}
\label{tab:External Open-Source Projects}
\begin{tabular}{|c|l|}
\hline
\textbf{Project} &
  \multicolumn{1}{c|}{\textbf{Description}} \\ \hline
Colosseum\text{\cite{ProjColo}} &
  \begin{tabular}[c]{@{}l@{}}This is a large-scale wireless testbed with open access and \\ public availability for research using virtualized and software waveforms. \\ It is hosted at the Northeastern University in Boston, USA \\ and the system is remotely accessible to users.\end{tabular} \\ \hline
OpenAirInterface\text {\cite{ProjOpenAirInt1}\cite{ProjOpenAirInt2}} &
  \begin{tabular}[c]{@{}l@{}}This project brings together a group of open-source software developers \\ who collaborateto create the RAN and CN technologies for wireless communication.\end{tabular} \\ \hline
srsRAN\text{\cite{ProjsrsRAN}} &
  \begin{tabular}[c]{@{}l@{}}An open source projected by Software Radio Systems (SRS) to develop a 5G software \\ radio suite.\end{tabular} \\ \hline
ONF\text {\cite{ProjONF}} &
  \begin{tabular}[c]{@{}l@{}}The Open Networking Foundation (ONF) is a non-profit organization that promotes \\ innovation in software-defined programmable networks. It is mainly operator-driven.\end{tabular} \\ \hline
ONAP\text {\cite{ProjONAP}} &
  \begin{tabular}[c]{@{}l@{}}ONAP is an open-source software platform that offers comprehensive \\ lifecycle management and real-time, policy-driven orchestration and \\ automation of physical and virtual network activities.\end{tabular} \\ \hline
Open Source MANO\text{\cite{ProjOpenMANO}} &
  \begin{tabular}[c]{@{}l@{}}In line with ETSI NFV models, ETSI OSM is aimed at creating an\\ open source Management and Orchestration (MANO) stack.\end{tabular} \\ \hline
O-RAN Gym\text {\cite{ProjGym}} &
  \begin{tabular}[c]{@{}l@{}}A publicly accessible research platform that enables large-scale data-driven O-RAN\\ experimentation. It provides an O-RAN compliant near-real-time RIC and E2 termination,\\ which can be used to build frameworks creation and testing of data-driven xApps.\end{tabular} \\ \hline
Open5GS\text {\cite{ProjOpen5GS}} &
  \begin{tabular}[c]{@{}l@{}}Open5GS is a C-language implementation of 5G Core and EPC, \\ i.e., the core network of NR/LTE network.\end{tabular} \\ \hline
Magma\text {\cite{ProjMagma}} &
  \begin{tabular}[c]{@{}l@{}}It is an open-source software platform that aims to provide network operators an \\ open and flexible mobile CN solution.\end{tabular} \\ \hline
\end{tabular}
\end{table}

\section{Open Issues and Future Research Directions}

Although O-RAN Alliance has been drafting specifications and collaborating with open source communities to enable open, interoperable, virtualized, and intelligent RAN, still there are issues and challenges, from network operators and standardization points of view, that need to be addressed to truly realize the features of Open RAN. In this section, we summarize the open issues, challenges, and future research directions that we identified in this study. 

\subsection{Open Issues and Challenges}

\begin{itemize}

\item \textbf{Architectural aspects:} The O-RAN reference architecture has evolved slowly over time since 2018. Meanwhile 3GPP has also been working continuously towards improving the 5G architecture and technology. A key requirement of O-RAN architecture is be as close as possible to 3GPP architecture. Hence, the specification work for O-RAN architecture is not fully completed and additional features, blocks, and functions would need to be introduced to enhance the capability of O-RAN. For instance, the O-RU termination of the O1 interface towards SMO and potential virtualization opportunities for the O-RU need to be studied and are earmarked as candidates for future studies [26]. Furthermore, the cooperation among multiple near-RT RICs and controlling them for data collection and execution (e.g., centralized, distributed, or hybrid) are important aspects that require further study. Similarly, communication between the near-RT RICs with 3rd party application servers and the possibilities of exposing the RAN capabilities need to be studied.
 
\item \textbf{Performance aspects:} The virtualization of O-RAN network functions offers a unique opportunity to improve RAN performance. Different O-RAN network functions and nodes can be migrated dynamically from one infrastructure to another (e.g., from regional cloud to edge cloud and vice versa) based on the load predictions or in-case of node/function failures. However, predicting such traffic congestion that might lead to overloading or predetermining any imminent failures and also arranging alternatives become a highly complex requirement especially in highly scaled and dynamic RAN deployments. Supporting time-sensitive use cases through resource migration or application migration without affecting the QoS and service continuity are some of the challenges that need to be addressed from a deployment and management perspective.

\item \textbf{Security aspects:} The O-RAN  Alliance WG11 has been studying the security aspects of the O-RAN architecture by considering the following aspects: thread modeling, risk assessment, security requirements, security mechanisms and protocols to meet the requirements, and security testing for validation and evaluation. However, only a few interfaces (e.g., Open FH) and network functions (near-RT RIC and non-RT RIC) related security aspects are studied \cite{W11f} \cite{W11e}. There is a lot of scope to explore and analyse the security aspects for various entities and interfaces. For instance, the security aspects of network functions (e.g., O-CU-CP, O-CU-UP, O-DU, O-RU, FHGW, and FHM) and interfaces (e.g., E2, R1, Y1, O-Cloud Notification, and Cooperative Transport Interface) need to be studied. Furthermore, the security aspects of shared cloud infrastructure, integration of nodes and network functions, open source software, secure lifecycle management of network functions also have to be studied.    
	
\item \textbf{AI/ML aspects:} The O-RAN Alliance prioritizes the usage of RICs to operate and maintain highly scaled network deployments, relying heavily on AI and ML based algorithms for training models. These are then used for extracting inferences through data analytics and transferring policy based guidelines to near-RT RIC via control loops to improve the performance of the RAN. However, the accuracy of the models and the validity of the inferred data is paramount to take the right decisions at the right times. This becomes a major challenge to overcome and there lies a potential of catastrophic impact in the case of safety and mission-critical applications if the accuracy and freshness of the inference data is not up to the expected level.

\item \textbf{Energy saving aspects:} Adapting AI and ML based techniques to support new services and applications, optimize network resources, and automate the RAN using RICs may consume more energy to train models as AI and ML techniques which employ deep neural networks often have large amounts of computational power requirement \cite{EI}, which may cause to global warming and climate change effects. Hence, the challenge of developing a sustainable and environment friendly advanced AI/ML techniques and energy-efficient architectures need to be considered.    
\end{itemize}

\subsection{Future Research Directions}

\begin{itemize}

\item \textbf{Blockchain:} With the advent of 5G networks and its diverse features, several new technologies such as SDN and NFV are being integrated into the 5G networks to fulfil the requirements \cite{Blockchain1}. However, integrating these technologies into the telecom network poses several challenges associated with decentralization, transparency, privacy, and security. To address these issues, blockchain technology has emerged as a viable solution due to its strengths such as  auditability, immutability and distributed architecture.
The O-RAN Alliance is specifically interested in developing a secure and interoperable RAN enabled by blockchain. The current O-RAN security relies on opt-in Public Key Infrastructure (PKI) based encryption and authentication solutions and transferring the trust to a centralized CA as a trusted party. This becomes a single point of failure for the entire network in the probability of a communication outage with the CA or a compromised CA due to malicious activities. To overcome such a vulnerable situation, security and authentication powered by a blockchain becomes appealing for network deployment \cite{Blockchain3}.
The O-RAN Alliance is working on a new type of RAN architecture called the Blockchain-enabled RAN (BE-RAN). The main focus of this architecture is supporting mutual authentication and blockchain-based PKI \cite{Blockchain2}. The potential areas of future research and analysis would include the impact of BE-RAN on existing security systems including zero trust system, certificate identity, and runtime security. Further studies would also be required to understand the impact and approach to adapting blockchain on the interfaces used for control plane, user plane, and synchronization plane \cite{Blockchain3}.

\item \textbf{Digital Twin:}
Digital Twins (DTs) are becoming an important part of the industrial manufacturing domain because it creates virtual simulations of physical assets such as factories, supply and transportation chains. Due to major advances in development of software platforms that leverage AI/ML methodologies and also availability of high-performance computing accelerated by GPUs, DTs have gained traction in many different areas, such as smart cities, manufacturing, and retail \cite{DigitalTwin3}.
With the advent of the network disaggregation paradigm as championed by O-RAN Alliance, the ensuing complexity of the networks and their management increases to such a degree that traditional deterministic engineering approaches might falter. Therefore, network simulation tools like Digital Twin Networks (DTNs) that run on a high-definition digital representation of the network become increasingly important. DTNs use real-time data and models to create an accurate simulation platform of the physical network, that can be used to provide up-to-date network status and also predict future network states \cite{DigitalTwin1}\cite{DigitalTwin2}. It also provides interfaces for communication with the physical network and other network applications/users.
The O-RAN Alliance's next Generation RG (nGRG) is working on ideas and use cases to leverage DTNs in 5G and 6G deployments. The main focus of future studies would be on defining requirements and design principles including reliability, latency, scalability, agility, and security \cite{DigitalTwin3}. The data exchange and the interfacing between DTNs and physical O-RAN components are also important areas that needs further research and analysis from an Open RAN perspective.

\item \textbf{Metaverse:} In one aspect, metaverse can be described as an immersive virtual world that is facilitated by the use of VR and AR headsets. In order to enable end users to connect into the metaverse via a 5G network, 3GPP has introduced the support for Extended Reality (XR which includes VR, AR, and MR) and cloud gaming devices as a subset of UEs. As part of Releases 16 and 17, 3GPP has identified some enhancements aimed at improving XR device latency and power efficiency. From an Open RAN perspective, further studies are required regarding on the optimal implementation aspects of these enhancements that include i) reduce bandwidth if no/low data, ii) switch to low power if no data through secondary cell dormancy, iii) cross-slot scheduling gap between control and data for sleep, iv) handle periodic traffic through uplink configured grant, v) switch to low power if no data through uplink skipping, and vi) faster transition to sleep after XR burst through control channel skipping \cite{Meta1}. 3GPP is planning for further research on various topics that can also be incorporated into the O-RAN architecture. These include i) low latency mobility using L1/L2 signalling for handoff, ii) staggering UE traffic arrival at gNB through improved scheduler, and iii) improve QoS based on multimedia traffic requirements and patterns.

\item \textbf{Non-Public/Private Networks:} A private 5G network, also termed as Non-Public Networks (NPN) by 3GPP, is a 5G network deployed for non-public use. As 5G network promises very high speed, ultra-high reliability, and low latency, 5G private network is a potential candidate for supporting non-public applications in many industry verticals such as smart manufacturing, Industry 4.0, transportation and logistics, airport, ports, mining, healthcare, education, and entertainment. 5G private networks can adapt O-RAN architecture to offer new services using RICs. However, secure operation, expanding coverage based on the UE mobility, sharing data and resources, and integrating edge cloud and exposing the capability are challenging and need further study.

\item \textbf{Non-Terrestrial Networks:} A Non-Terrestrial Network (NTN) refers to a network for communication purposes which partially or fully operates through a space-borne vehicle, i.e., using satellites in Geostationary Earth Orbit, Medium Earth Orbit, and Low Earth Orbit, or an airborne vehicle  (e.g., High Altitude Platforms and Unmanned Aerial Vehicles). The most important feature that makes NTNs unique is their capability to provide connectivity in unreachable areas (i.e., ocean vessels and airplanes) or remote areas (i.e., rural areas) where huge investment is required to build a terrestrial infrastructure. 3GPP has introduced new reference architectures to implement NTNs\cite{3GPP_38.821} as part of Release 15. These are i) NG-RAN architecture with transparent satellite and ii) NG-RAN architecture with regenerative satellite. New components such as NTN-Gateway, NTN-Payload, NTN feeder link, NTN service link, and NTN control functions have been introduced. Future studies should focus on the approach to standardize these newly introduced components and interfaces in an open manner.

\item \textbf{Precise Positioning and Ranging:} In order to implement highly precise positioning of UEs, 3GPP has introduced a new function called Location Management Function (LMF). The LMF receives positioning measurements and information from the NG-RAN and UEs, via the Access and Mobility Management Function (AMF) \cite{LMF1}. A new interface called NLs interface is introduced in the 5G core for communication between AMF and LMF. Additionally, a new NR Positioning Protocol A (NRPPa) is introduced to carry the positioning information from the RAN to the LMF over NG-C interface. The LMF can also configure the UE using the LTE Positioning Protocol (LPP) via AMF and through the gNB. In order to support precise positioning and ranging functionalities in O-RAN, further studies are required regarding the optimal implementation aspects of NRPPa and LPP protocols. Also further analysis would be required on how the measurements data is collected from the O-RAN components and carried via the NG-C interface.

\item \textbf{Explainable AI:} In recent years, Deep Learning (DL) based techniques are employed to successfully handle complex and hard tasks that are beyond the limit of human operators. DL based techniques can be applied for training models in non-RT RIC for specific RAN functionalities and the trained models can be shared with near-RT RIC for real-time operations. However, DL based models are considered as black-box and thus it is difficult to understand the underlying operations and reasons that why the model have taken certain complex actions and decisions \cite{XAI}. This lack of transparency may lead to an issue of vulnerability to attacks and inject malicious data. To offer seamless and secure services, it is expected that AI/ML techniques should be explainable, robust, and verifiable. To handle this issue, explainable AI and adversarial AI techniques can be explored.  

\end{itemize}

\section{Conclusion}

The Open Radio Access Network (Open RAN) is a new paradigm which revolutionizes the telecom industry to move towards open, interoperable, virtualized, and intelligent RAN, while improving the performance, agility, and cost efficiency. It enables openness through open interfaces and running network functions on whitebox hardware by leveraging the open source modular software.  In this paper, we first presented a comprehensive overview of evolution of RAN and the O-RAN Alliance standardization activities. We also discussed about security aspects of O-RAN architecture, use cases, deployment aspects, and relevant open source projects. Finally, we summarized the open issues, challenges, and future research directions for future study. We hope this work can serve as a good reference to understand the O-RAN standardization activities and pursue further research in this field to support and realize the features of Open RAN movement.

\section*{acknowledgments}
We would like to thank Ritesh Kumar Kalle for his comments on the paper for further improvements. We also thank Balaji Durai, Manikantan Srinivasan, and Murugesan P for initial discussions to form a team.

\bibliographystyle{IEEEtran}
\bibliography{M1_ref}

\end{document}